%% file: main.tex
\documentclass[format=acmlarge, review=false]{acmart}
\usepackage{acm-ec-20-modified}
\usepackage{booktabs} 
\usepackage[ruled]{algorithm2e} 

\SetAlFnt{\small}
\SetAlCapFnt{\small}
\SetAlCapNameFnt{\small}
\SetAlCapHSkip{0pt}
\IncMargin{-\parindent}

\setcitestyle{acmnumeric}

\input{DONG_ROY-macros}

\title[Protecting Consumers Against Personalized Pricing]{Protecting Consumers Against Personalized Pricing:\\A Stopping Time Approach}

\author{Roy Dong, Erik Miehling, Cedric Langbort}

\begin{abstract}
The widespread availability of behavioral data has led to the development of data-driven personalized pricing algorithms: sellers attempt to maximize their revenue by estimating the consumer's willingness-to-pay and pricing accordingly. Our objective is to develop algorithms that protect consumer interests against personalized pricing schemes. In this paper, we consider a consumer who learns more and more about a potential purchase across time, while simultaneously revealing more and more information about herself to a potential seller. We formalize a strategic consumer's purchasing decision when interacting with a seller who uses personalized pricing algorithms, and contextualize this problem among the existing literature in optimal stopping time theory and computational finance. We provide an algorithm that consumers can use to protect their own interests against personalized pricing algorithms. This algorithmic stopping method uses sample paths to train estimates of the optimal stopping time. To the best of our knowledge, this is one of the first works that provides computational methods for the consumer to maximize her utility when decision making under surveillance. We demonstrate the efficacy of the algorithmic stopping method using a numerical simulation, where the seller uses a Kalman filter to approximate the consumer's valuation and sets prices based on myopic expected revenue maximization. Compared to a myopic purchasing strategy, we demonstrate increased payoffs for the consumer in expectation.
\end{abstract}

\begin{document}

\maketitle


\section{Introduction}
\label{sec:intro}

\input{intro}

The rest of this paper is organized as follows. 
In Section~\ref{sec:background}, we present the literature most related to our work and contextualize our contributions. 
In Section~\ref{sec:model}, we outline the mathematical model of the consumer's decision, and define her optimization problem. 
With the model introduced, we characterize the consumer's optimal decisions in Section~\ref{sec:theory}. 
We present an algorithm for the computation of the consumer's decisions in Section~\ref{sec:computation}, which approximates the optimal stopping time.
In Section~\ref{sec:example}, we present a numerical example demonstrating how an algorithmic stopping time approach can improve the consumer's payoff in a stylized model of surveillance capitalism. Finally, we close with some concluding remarks in Section~\ref{sec:conclusion}.

\subsection{Notation}

Throughout this paper, we use the following notation. For a vector $v$ with index set $I$, we will frequently write $v$ in terms of its elements $v = (v_i)_{i \in I}$ or, in brief, $(v_i)_i$. All random elements will be with respect to the same underlying probability space $(\Omega, \mc{A}, P)$. We write $\mb{E}$ to denote the expectation operator, and may subscript $\mb{E}_{X \sim \nu} f(X)$ to denote that $X$ is distributed according to the probability measure $\nu$. For any sub-$\sigma$-algebra $\mc{F} \subseteq \mc{A}$, we write $Y \in \mc{F}$ to denote that $Y$ is a $\mc{F}$ measurable function. We represent the conditional expectation operator as $\mb{E}[\cdot \vert \mc{F}]$; this is the projection operator in $L^2$ onto the space of $\mc{F}$ measurable functions, i.e. for any random variable $X$, we define $\mb{E}[X \vert \mc{F}]$ as the $P$ almost surely unique $\mc{F}$ measurable function such that $\mb{E}[\mb{E}[X \vert \mc{F}] Y] = \mb{E}[X Y]$ for any $Y \in \mc{F}$. For a random element $X$, we write $\sigma(X)$ to denote the $\sigma$-algebra generated by $X$. We let $f \vee g$ denote the point-wise maximum of two functions $f$ and $g$. 


\section{Related Literature}
\label{sec:background}

\input{back}


\section{Model}
\label{sec:model}

\input{model}


\section{Discrete-Time Optimal Stopping Theory}
\label{sec:theory}

\input{calc}
\section{Computation of Optimal Stopping Times}
\label{sec:computation}

\input{computation}


\section{Numerical results}
\label{sec:example}

\input{example}


\section{Conclusion}
\label{sec:conclusion}

\input{conclusion}


%
%
%
%
%


\bibliographystyle{ACM-Reference-Format}
\bibliography{./DONG_ROY-refs}




\end{document}

%% file: DONG_ROY-macros.tex
\usepackage{amsmath}
\usepackage{amssymb}
\usepackage{amsthm}

\newcommand{\eps}{\epsilon}

\newcommand{\mc}{\mathcal}
\newcommand{\mb}{\mathbb}

\newtheorem{definition}{Definition}

\newtheorem{proposition}{Proposition}

\newtheorem{remark}{Remark}


%% file: intro.tex

The widespread availability of behavioral data has led to the development of data-driven personalized pricing algorithms: sellers attempt to maximize their revenue by estimating the consumer's willingness-to-pay and pricing accordingly. Thus, if some observable features about you suggest that you may value a product more, you will see a higher price from sellers using these pricing algorithms. In the current age of data collection, there are many observable features for each individual, and these algorithms have increasingly refined methods to target customers and price to extract as much consumer surplus as possible.

Whether or not these methods are beneficial to society, or ethical, still remains an open question. One could argue that personalized pricing allows more voluntary trade to occur than would be possible with uniform pricing, and thus the societal benefits from trade are increased as a result of personalized pricing. In this light, personalized pricing moves the market to a more efficient allocation~\cite{Varian:1985aa}. However, there is also evidence that many of these algorithms aversely affect certain populations more than others~\cite{Barocas:2019aa}: race and gender are, in particularly, commonly used for statistical inference about reservation prices.

Regardless of the societal or ethical implications, personalized pricing schemes are currently deployed in practice. 
Airlines use them to price ancillary services and products, such as in-flight meals, baggage allowance, and legroom~\cite{Shukla:2019aa}. On websites such as Orbitz, Expedia, Hotels.com, and Priceline, the offered options and pricing of plan tickets, hotels, and car rentals depend on the consumer's features (e.g. OS/browser, age, income) and browsing history, which are assumed to be correlated with their valuation of such goods~\cite{Hannak:2014aa}. Web-pages use tracking to determine how much to charge advertisers to place their ads~\cite{Golrezaei:2019aa}. Similarly, e-retailers will set prices based on consumer context~\cite{Hannak:2014aa}. For example, Home Depot and Staples were both known for increasing prices if a consumer's zip code was farther from a brick-and-mortar store. 

Our objective is to develop algorithms that protect consumer interests against personalized pricing schemes. In this paper, we consider a consumer who learns more and more about a potential purchase across time, while simultaneously revealing more and more information about herself to a potential seller. For example, a consumer may browse the website to learn more about airline tickets or a pair of boots, and, as she leaves behind a trail of cookies across multiple websites, retailers increasingly have a better sense of her reservation prices and can set prices accordingly. It is unclear how a consumer should behave in these settings. Should she purchase earlier, at a higher level of uncertainty, to reduce the information she reveals to sellers? Should she take a risk and wait for prices to drop in the future, even though a purchase at the current moment would yield positive utility? 

The main contributions of this work can be summarized as follows.
\begin{itemize}
	\item We formalize a strategic consumer's purchasing decision when interacting with a seller who uses personalized pricing algorithms, and contextualize this problem among the existing literature in optimal stopping time theory and computational finance.
	\item We provide an algorithm that consumers can use to protect their own interests against personalized pricing algorithms. This algorithmic stopping method uses sample paths to train estimates of the optimal stopping time. To the best of our knowledge, this is one of the first works that provides computational methods for the consumer to maximize her utility when decision making under surveillance.
	\item We demonstrate the efficacy of this algorithm using a numerical simulation, where the seller uses a Kalman filter to approximate the consumer's valuation and sets prices based on myopic expected revenue maximization. Compared to a myopic purchasing strategy, we demonstrate increased payoffs for the consumer in expectation.
\end{itemize}

%% file: back.tex

In this section, we provide an overview of the existing literature related to this work. 

Personalized pricing is a special form of dynamic pricing. Dynamic pricing, broadly speaking, refers to any pricing methods where the seller uses his available information to set prices. For example, in the 1990s, Coca-Cola researched how to make vending machines whose prices varied with temperature: on a hot day, charge more for a can of soda. More recently, thanks to the widespread proliferation of data collection, personalized pricing has become feasible: these methods set prices for an individual based on their features. 
For an overview of the research on dynamic pricing, we refer the reader to the survey by~\citet{Boer:2015aa}.

The work closest in spirit to ours is in the research of computation and theory for personalized pricing methods. 
Broadly speaking, the related work can be divided into two categories: 1) work that focuses on algorithmic methods to compute pricing strategies using available data, and 2) work that focuses on characterizing game-theoretic equilibrium of the interactions between a consumer and a seller when the seller uses personalized pricing strategies.

Typically, from an algorithmic perspective, one supposes that there is an unknown mapping between the consumer's observable features and their willingness-to-pay, and the goal of these methods is to learn this mapping online while simultaneously experiencing rewards/costs during the learning process. The observable features are often referred to as contextual information, and include data such as a consumer's browsing history.

It is not uncommon to assume that the consumers are myopic or non-strategic. In this case, the seller sets prices to trade-off exploration and exploitation: exploration seeks to learn how observable features affect the consumer's reservation prices, while exploitation seeks to use existing knowledge to maximize revenue in the short term. \citet{Cohen:2016aa} and~\citet{Mao:2018aa} study the problem of dynamic pricing for a set of differentiated products, and how observable features affect the market value of these products. \citet{Qiang:2016aa} learns a demand function for a product and maximizes reward, assuming that some demand covariates are observed at every time step. \citet{Javanmard:2019aa} explore dynamic pricing techniques when the features live in a high-dimensional space. \citet{Nambiar:2018aa} use random price shocks to encourage exploration. \citet{Ban:2019aa} models personalized pricing with repeated interactions with a consumer, but this consumer's demand is a fixed function of the price offered and the observable features. \citet{Shah:2019aa} studies the problem of revenue maximization in a semi-parametric setting, where the success/failure of each sale is determined by if the price is below the consumer's valuation. 
\citet{Levina:2009aa} uses an online learning approach to maximize revenue accounting for time, remaining inventory and price, viewing the consumer's behavior as a fixed and known function of time. 
Effectively, in all of these works, the formulation is equivalent to a non-strategic consumer who makes a purchase when their product valuation exceeds the offered price.

More relevant to our approach is work that assumes the consumers are strategic. Some applications of dynamic pricing do not permit the assumption that consumers are myopic. In this case, issues of incentive compatibility and robustness to strategic manipulation become the focus of the pricing methods. 
\citet{Amin:2013aa} model the situation where a seller uses repeated interactions with a consumer to learn about the consumer's underlying valuation. In this setting, the consumer knows her valuation and is strategic. Under the assumption that the seller is more patient than the consumer (i.e. the consumer discounts the future more than the seller), \citet{Amin:2013aa} provide a no-regret algorithm, where regret is defined against the revenue when the consumer is non-strategic.
\citet{Kanoria:2014aa} assume the seller uses historical information to infer the distribution of valuations, which opens the seller up to strategic manipulation. They demonstrate how to implement an incentive-compatible mechanism. 
\citet{Golrezaei:2019aa} consider repeated interactions between the same seller and consumer, and they assume the consumer is strategic. They develop pricing algorithms for the seller that utilize contextual information and are robust to the strategic decisions of the consumer. They introduce epochs which introduce a lag before a consumer's data is used to determine prices; because of the consumer's discount factor, this lag reduces the gains from strategic manipulation. Additionally, the seller periodically clears his data, which limits the long-term benefits for the consumer from strategically manipulating data. Under certain conditions, their method ensures that consumers cannot gain from strategically manipulating their data. These works focus on developing algorithms for the seller, 

In contrast to these works, we focus on how consumers can react to personalized pricing algorithms. We view the pricing strategies of the seller as both fixed and given, and define how the consumer can behave to maximize their own payoff, subject to their own informational constraints. Against robust algorithms or incentive-compatible mechanisms, the method in this paper may not provide much benefit for the consumer. However, many of the personalized pricing algorithms in practice are not designed to handle strategic behavior, and, in these settings, our proposed methods can provide the consumer with some utility gain.

Complementing the algorithmic approaches to this problem are more game-theoretic approaches. These methods seek to characterize the equilibrium behavior of agents. 
\citet{Stokey:1979aa} considers as game where the seller announces prices in advance, and the consumer must decide when to buy or reject the good.
\citet{Besanko:1990aa} characterize the sub-game perfect Nash equilibrium in seller/consumer interactions using dynamic programming. Their model features hidden information which slowly gets revealed by the agent's actions across time. 
\citet{Chade:2002aa} consider an infinite-horizon, two-type model, and characterize how the seller's best responses change as information revelation varies.
\citet{Su:2007aa} supposes that the consumer takes on four possible types, which correspond to combinations of high/low valuation and patient/impatient discount factors, and characterize how the equilibrium change as the distribution of types vary.
\citet{Loginova:2008aa} consider a two-period model and demonstrate when price discrimination would occur. 
\citet{Aviv:2008aa} analyze how to optimally price a good when the value of the good is time-varying: prices are declared in advance, they analyze the effect of strategic consumers on expected revenues. In this work, the strategic consumer's strategy is shown to be a thresholding policy.
\citet{Levin:2010ab} prove the optimality conditions for a sub-game perfect equilibrium for the stochastic dynamic game between the consumer and the seller, also utilizing a dynamic programming approach. 

The main difference in our work is that we consider the process of exogenous information discovery. 
Prior work in these game-theoretic formulations focused on the effects of {signaling}: in particular, the consumer knows her own valuation, and when she does not purchase the good at time $t$ for price $p_t$, she reveals information about her private valuation. 
In our work, we focus on the effects of information discovery: the consumer does not know her own valuation, and as time passes, the consumer learns more information about her own valuation. Meanwhile, the seller learns more contextual information about the consumer. Practically, this difference means that their formulations admit a dynamic programming analysis different from ours: we must use conditional expectations and the Snell envelope due to how the consumer learns about her valuation across time. In contrast, the previous works could implement a value function directly on the action space.

Additionally, due to the complexity of these processes, in our work we restrict ourselves to fixing the seller's pricing strategies and computing the consumer's approximate best response. This contrasts with the cited works, which either characterize equilibrium of the strategic interactions between the consumer and seller or characterize the seller's best response. Thus, our focus in this paper is providing computational methods for the consumer, which can be applied in any settings where a consumer must learn under surveillance.

Finally, much of our computational methods are drawn from the mathematical finance literature. The problem of computing an optimal stopping time arises in the pricing of American options, which are options that can be exercised at any time. The value of an American option must weigh the value of being able to exercise the option in the future with the known payoff of exercising the option immediately. For an overview of the financial interpretations of optimal stopping times, we refer the reader to~\citet{Follmer:2008aa}. 
In this paper, we use Monte Carlo methods for the computation of stopping times, which were developed in part by~\citet{Carriere:1996aa} and~\citet{Tsitsiklis:1999aa} but popularized in the computational finance community by~\citet{Longstaff:2001aa}. 
For a theoretical analysis of pricing American options, we refer the reader to~\citet{Haugh:2004aa} and~\citet{Rogers:2002aa}.

%% file: model.tex

In this section, we introduce our model for the consumer's purchasing decision. 
We define the relevant payoffs when the consumer makes a purchase at time $t$, when she stops browsing without a purchase at time $t$, and when she chooses to continue browsing at time $t$. The setting is dynamic, so we specify our model for the consumer's state of information and its evolution, as well as the relationship between the payoffs of the consumer at different steps in time. 


\subsection{Overview}

Our model is a discrete-time, finite horizon interaction, with time indices $\mb{T} = \{0,1,\dots,T\}$. 
We assume that the consumer is considering the purchase of a single, indivisible good. To decide whether or not to do so, she learns about her valuation at each time step, while simultaneously revealing information to a seller. Throughout this document, we will refer to the `consumer' with the pronouns she/her and, similarly, we will refer to the `seller' with the pronouns he/him.

As is common in the literature~\cite{Aviv:2008aa,Golrezaei:2019aa}, we assume that the seller credibly commits to pricing strategies $(p_t)_{t \in \mb{T}}$ prior to the consumer's first decision at $t = 0$. Here, $p_t$ are random variables, which incorporate the seller's information about the consumer. 
This assumption is motivated by the general observation that when consumers are fully informed and rational, firms earn more revenue when they can credibly pre-commit to a pricing strategy~\cite{Hart:1988aa,Salant:1989aa,Aviv:2008aa}. 

The order of events at each time step $t \in \mb{T}$ are:
\begin{itemize}
\item Given his available information at $t$, the seller calculates the offered price at time $t$, denoted $p_t$, and communicates this price to the consumer.
\item The consumer sees the price and then decides between 3 choices: 1) purchase the good, receive a payoff $\pi_t$ that depends on her current information and price, and end the interaction; 2) reject the good, receive a payoff of $0$, and end the interaction; 3) continue browsing, and the game continues. If $t = T$, then choice (3) is unavailable, and the consumer must either purchase or reject the good.
\item If the consumer chooses to continue, both the consumer and the seller learn of new information, and the process repeats.
\end{itemize}


\subsection{Consumer model}
\label{sec:consumer_obj}

In this subsection, we present the formal details of our model. 
As stated previously, the underlying probability space is denoted $(\Omega, \mc{A}, P)$. 

The seller chooses prices based on his available information; these prices are modeled as random variables. $p_t(\omega)$ denotes the price offered at time $t$, and we will refer to $p_t : \Omega \rightarrow \mb{R}$ as a {\it pricing strategy}. The offered price $p_t(\omega)$ implicitly depends on the seller's information about the consumer.
The consumer's goal is to decide when to purchase or reject the good in an optimal way, based on her available information at each time step. Since the seller announces his pricing strategies at the beginning of the game, we will consider the functions $(p_t)_{t \in \mb{T}}$ as fixed and known.

We represent the consumer's {\it state of information} at each time $t$ as a $\sigma$-algebra: let $\mc{F}_t \subseteq \mc{A}$ denote the consumer's state of information at time $t$. 
We assume that the consumer has perfect recall, i.e. she remembers all past information available to her. Formally, this means that the sequence $\mc{F} = (\mc{F}_0,\mc{F}_1,\dots,\mc{F}_T)$ is a {filtration}; in other words, $\mc{F}_0 \subseteq \mc{F}_1 \subseteq \dots \subseteq \mc{F}_T$. Since the price $p_t(\omega)$ is announced at time $t$, we note that $p_t(\omega)$ is known to the consumer at time $t$, i.e. $p_t \in \mc{F}_t$. For clarity of presentation, in the sequel we well drop the $\omega$ argument from random variables.

First, we define the consumer's payoffs when she purchases the good. In our model, if she purchases the good at time $t$, she will receive payoff $\pi_t$, where $\pi_t$ is an ${\mc{F}_t}$ measurable function which depends on the offered price $p_t$. That is, we assume she knows what payoff she will receive if she chooses to purchase the good at time $t$. 

\begin{remark}
\label{remark:not_know}
We emphasize that knowing $\pi_t$ is {\em not} the same as assuming that she knows her valuation of the good. We simply assume that she knows the payoff she will get from purchasing the good, at her current state of information's risk and uncertainty. 
\end{remark}

%

On the other hand, if she rejects the good at time $t$, she will neither receive the good nor pay any price. In this case, she receives payoff $0$. 

With her payoffs defined, we note that we can combine the `purchase' and `reject' actions into a single `exit' action. In this case, she ends the game and receives payoff $H_t = \pi_t \vee 0$. This means that whenever she chooses to `exit' and end the game, she will always choose the higher payoff between `purchase' and `reject'. Much of the analysis from this point on is greatly simplified notationally by the introduction of this `exit' action with payoff $H_t$. 


Now, we introduce the notion of a stopping time for discrete-time filtrations.

\begin{definition}
For any filtration $\mc{F}$, we say that a $\mb{T}$-valued random element $\tau$ is a {\em stopping time adapted to $\mc{F}$}, or sometimes a {\em $\mc{F}$ stopping time}, if the set $\{ \tau = t \}$ is $\mc{F}_t$ measurable for all $t \in \mb{T}$.
\end{definition}

Thus, the consumer's decision is to pick a stopping time, i.e. she decides whether or not to `exit' based on the information currently available to her. Given a stopping time, the choice of whether to `purchase' or `reject' when exiting is straightforward: whichever has the higher payoff will be chosen. When the consumer chooses a stopping time $\tau$, the payoff she receives in expectation is $\mb{E}[H_{\tau}]$.

In summary, the consumer's decision of when to `purchase' or `reject' the good can be reduced to a choice of when to `exit', given her available information. The payoff from `exit' (denoted $H_t$) is the maximum of the payoff from `purchase' (denoted $\pi_t$) and the payoff from `reject' (which is $0$). When the seller's pricing strategies $(p_t)_t$ are treated as fixed, the processes $(\pi_t)_t$ and $(H_t)_t$ can be taken as given. Her feasible strategies of when to `exit' (given her available information) is formalized as a stopping time, and the ex-ante expected payoff she receives is $\mb{E}[H_{\tau}]$.

\subsection{Example: Gaussian model with myopic pricing}
\label{sec:example_model}

In this subsection, we outline a specific example of the general model outlined above. 
We consider the case where all the relevant values and observations are drawn from Gaussian distributions. In this model, the consumer learns about her true underlying valuation across time. Meanwhile, the seller receives noisy measurements of the consumer's valuation, which, at each time step, allows the seller to better estimate the true underlying valuation as well. Additionally, the seller sets prices to myopically maximize his expected revenue at each time step. This example will be used for numerical simulations in Section~\ref{sec:example}.

We model the consumer's valuation process as follows. Initially, the consumer has some initial estimate of her valuation, $v_0$. At each time $t$, she makes an observation $\eps_t$, where $\eps_t > 0$ increases her valuation of the good and $\eps_t < 0$ decreases her valuation. More formally, she learns $v_t = v_{t-1} + \eps_t$, where $\eps_t \sim N(0,\sigma_\eps^2)$ independently across time. Her final valuation is denoted $v = v_T$, and is interpreted as the true underlying valuation. Thus, her posterior of $v$ given her available information at time $t$ is $v \sim N(v_t,(T-t)\sigma_\eps^2)$. As time progresses, the variance is decreasing, until time $t = T$ when she knows her true valuation $v$ with certainty. 

In addition, the consumer is offered a price $p_t$ at time $t$. This means that, should the consumer accept the price and make a purchase, her rewards are distributed according a Gaussian random variable with mean $v_t - p_t$ and variance $(T-t) \sigma_{\eps}^2$. We assume the consumer is risk averse, and, in particular, her utility can be modeled by a CARA (constant absolute risk aversion) utility function, given by $u(x) = 1 - \exp(-\gamma x)$ for some parameter $\gamma > 0$. Thus, at time $t$, should she make a purchase, her utility from purchase will be $\pi_t = 1 - \mb{E}[\exp(-\gamma X)]$ where $X \sim N(v_t - p_t, (T-t)\sigma_{\eps}^2)$. This can be calculated from the moment generating function of a normal distribution, which yields the utility:
\begin{equation}
\label{eq:pi_def_gauss}
\pi_t = 1 - \exp(- \gamma (v_t - p_t)) \exp\left( \frac{1}{2} \gamma^2 (T-t)\sigma_{\eps}^2\right)
\end{equation}
We note that even though $(v_t)_t$ is a martingale and the valuation remains the same in expectation across time, ceteris paribus, the consumer's payoff increases across time. This is due to the risk aversion in her utility function: holding other factors constant, as time increases, her uncertainty decreases (i.e. the variance of $v$ given $v_t$ is decreasing across time), which leads to higher utility.

In this example, her state of information at time $t$ is $\mc{F}_t = \sigma(v_0,\eps_1,\dots,\eps_t,p_0,\dots,p_t)$. We can see that $\pi_t$, defined in Equation~\eqref{eq:pi_def_gauss}, is $\mc{F}_t$ measurable. As mentioned in Remark~\ref{remark:not_know}, even though the consumer does not know $v = v_T$ at time $t < T$, she can calculate $\pi_t$, which is her payoff from a purchase. The payoff from exit is defined as $H_t = \pi_t \vee 0$, and any stopping time $\tau$ adapted to the filtration $\mc{F} = (\mc{F}_0,\dots,\mc{F}_T)$ will yield expected payoff $\mb{E}[H_\tau]$.

Now, we define the seller's pricing strategies $(p_t)_t$. 
From the seller's perspective, we suppose he receives noisy observations of the consumer's current valuation $v_t$. Thus, at time $t > 0$, the seller receives $y_t = v_t + \xi_t$ where $\xi_t \sim N(0,\sigma_{\xi}^2)$ are independent across time. The seller also has a prior on the consumer's initial valuation, so, from his perspective, at time $t = 0$, $v_0 \sim N(\mu,\sigma_v^2)$ for some parameters $\mu, \sigma_v^2$. Note that all the measurements of the seller are actually functions of information available to the consumer, i.e. the seller has less information than the consumer about her valuation. 
As such, the prices do not inform the consumer about her valuation.

In this formulation, the seller's estimation problem is a linear-quadratic estimation problem. That is:
\[
v_0 \sim N(\mu,\sigma_v^2) \qquad v_{t+1} = v_t + \eps_t \qquad \eps_t \sim N(0,\sigma_\eps^2) \qquad y_t = v_t + \xi_t \qquad \xi_t \sim N(0,\sigma_\xi^2)
\]
The posterior of $v_t$ given $y_0,\dots,y_t$ is a Gaussian random variable, whose mean and variance can be calculated recursively using the Kalman filter.

The seller uses his estimate of the consumer's valuation to set prices. 
At time $t$, we suppose the seller chooses $p_t$ that maximizes $p \cdot \Pr(v_t > p)$. This assumes two things. First, the seller myopically optimizes the expected revenue in their current time step. 
Second, he believes the consumer is similarly myopic; he believes the consumer will purchase the good whenever their valuation $v_t$ exceeds the offered price $p_t$, i.e. whenever $v_t > p_t$. 

So, the seller's behavior at each time step is as follows. He calculates the posterior of $v_t$ given $y_0, \dots, y_t$ using the Kalman filter, and then he chooses $p$ to maximize $p \cdot \Pr(v_t > p)$, which determines the price $p_t$ offered at time $t$. Noting the distribution of $v_t$, this amounts to maximizing $p \cdot Q\left(\frac{p - \mu_t}{\sigma_t}\right)$, where $\mu_t$ and $\sigma_t^2$ are the mean and variance of the seller's posterior at time $t$ (calculated by the Kalman filter), and the $Q$-function is $Q(z) = \Pr(Z > z)$ for a standard normal random variable $Z$.

%% file: calc.tex

In Section~\ref{sec:model}, we introduced our mathematical formulation of the consumer's decision of when to purchase or reject a good, accounting for her incomplete and dynamically evolving information state. Her decision reduces to the choice of a stopping time, and we defined the ex-ante expected payoff she receives for any given stopping time $\tau$. In this section, we will introduce the formalism required to define an optimal stopping time, and provide the theoretical tools to calculate the optimal stopping time. However, the theoretical tools may not be easy to compute in practice, which we address in Section~\ref{sec:computation}.


\subsection{Optimal stopping times}

To define an optimal stopping time, we first must introduce the definition of the essential supremum of a family of random variables. Note that this definition differs from the more common notion of essential supremum, which refers to the essential supremum of a single random variable.

\begin{proposition}[Existence of the essential supremum (\cite{Follmer:2008aa}, Theorem A.32)]
\label{prop:ess_sup_exist}
For any arbitrary collection of random variables $\Phi$, there exists a countable subset $\Phi_c \subseteq \Phi$ with the following property: if we define $\varphi^*(\omega) = \sup_{\varphi \in \Phi_c} \varphi(\omega)$, then $\varphi^* \ge \varphi$ almost surely for all $\phi \in \Phi$. 

Additionally, this random variable $\varphi^*$ is almost surely unique in the sense that for any random variable $\psi$, the following holds: $\psi \ge \varphi$ almost surely for all $\varphi \in \Phi$ implies that $\psi \ge \varphi^*$ almost surely.
\end{proposition}

\begin{definition}[Essential supremum]
We call $\varphi^*$, as defined in Proposition~\ref{prop:ess_sup_exist}, the {\em essential supremum} of $\Phi$, and denote it $\mathrm{ess}\sup~\Phi$. 
\end{definition}

The definition of essential supremum is needed due to technical conditions, as the point-wise supremum over an arbitrary set of random variables need not be measurable. The essential supremum provides a measurable function $\varphi^*$ which almost surely satisfies the desired properties of a point-wise supremum.\footnote{We can see $\varphi^*$ is measurable as it is defined as the supremum over a {\it countable} set of random variables.}

Let $\mc{T}$ denote the set of stopping times adapted to ${\mc{F}}$. 
An {\it optimal stopping time} $\tau^* \in \mc{T}$ is one that satisfies the equality almost surely:
\[
\mb{E}[H_{\tau^*}] = \mathrm{ess}\sup_{\tau \in \mc{T}}~\mb{E}[H_{\tau}]
\]


\subsection{The Snell envelope}

Our formulation allows us to cast the consumer's decision as an optimal stopping time problem, which allows us to solve for the optimal purchasing strategy using dynamic programming approaches. For the discrete-time, finite-horizon problem, the Snell envelope provides a method to calculate the optimal stopping time. 


Let us define the Snell envelope of the random process $H$.
\begin{definition}[Snell envelope]
Let $H$ be a random process adapted to $\mc{F}$. The {\em Snell envelope $U$ of $H$} is the random process recursively defined as follows. $U_T = H_T$, and $U_t = H_t \vee \mb{E}[U_{t+1}\vert\mc{F}_t]$ for $t < T$.
\end{definition}
From this definition, we make a few quick observations. 
First, $U$ is adapted to $\mc{F}$. Second, $U$ is a supermartingale by construction: $U_t \ge \mb{E}[U_{t+1}\vert\mc{F}_t]$ for each $t < T$. Third, $U$ dominates $H$, again by construction: $U_t \ge H_t$. 

While the process $H$ represents the value $H_t$ of `exit' at time $t$, the Snell envelope $U_t$ at time $t$ represents the value of having the flexibility of `exiting' at any time $s \in \{t,\dots,T\}$, where the payoff from `exit' at each time $s$ is $H_s$. At time $T$, the consumer {\em must} `exit', so $U_T = H_T$. Prior to that, $U_t$ must always at least be worth the value of `exit' the option at time $t$, and its value is the maximum of the current value of `exit' $H_t$ and the expected value of optimally choosing when to `exit' from time $t+1$ onward, conditioned on the currently available information $\mb{E}[U_{t+1}\vert\mc{F}_t]$.

We define the stopping time $\tau_{\min}$ as follows:
\[
\tau_{\min} = \min~\{s \in \mb{T} : U_s = H_s\}
\]
Note that since $U_T = H_T$, the minimum is well-defined, and since both terms are $\mc{F}$ adapted, $\tau_{\min}$ is a valid stopping time.

\begin{proposition}[Optimality of $\tau_{\min}$ (\cite{Follmer:2008aa}, Theorem 6.20)]
\label{prop:snell_solution}
For the Snell envelope $U$ of $H$, we have 
$U_0 = \mb{E}\left[H_{\tau_{\min}}\right] = \mathrm{ess}\sup_{\tau \in \mc{T}} \mb{E}[H_\tau]$.
\end{proposition}

Proposition~\ref{prop:snell_solution} tells us that we can use the Snell envelope to explicitly calculate the consumer's best strategy against a seller who credibly commits to pricing strategies $(p_t)_t$ in advance.
That is, at each time step $t$, it is optimal for the consumer to `exit' if $\tau_{\min} = t$.

%% file: computation.tex

In Section~\ref{sec:theory}, we outlined how the consumer's optimal stopping time problem can be solved using the Snell envelope. This provides us with an algorithm which can maximize a consumer's welfare in settings of surveillance capitalism, where a consumer must trade-off between learning more about a product and revealing more about her preferences. The main modeling assumption that allowed us to use optimal stopping time methods was the assumption that the seller pre-commits to pricing strategies. This assumption allows us to take the pricing strategies $(p_t)_t$ as given, which removes many of the game-theoretic interactions between the consumer and the seller.

However, even in this simplified case, the algorithm provided in Section~\ref{sec:theory} is primarily theoretical. Modeling the filtration $\mc{F} = (\mc{F}_t)_t$ for realistic pricing functions is non-trivial, and, furthermore, calculating the appropriate conditional expectations and recursions may be computationally quite difficult. In this section, we present an algorithm which allows the consumer to explicitly calculate an approximation of her optimal stopping time. 

Throughout this section, we assume that the consumer has the capacity to sample from her valuations and the (potentially random) pricing strategy. In particular, she can draw independent path samples $(H_t^n)_{t \in \mb{T}}$ to train an estimator for the Snell envelope. 


\subsection{Algorithmic stopping method}

In this paper, we use Monte Carlo methods to estimate the conditional expectation operators. As mentioned in Section~\ref{sec:background}, the method presented here builds on the works of~\citet{Carriere:1996aa},~\citet{Tsitsiklis:1999aa}, and~\citet{Longstaff:2001aa}. The algorithm takes $N$ sample paths $(H_0^n,\dots,H_T^n)_{n \in N}$ as input. From this, the algorithm proceeds backwards in time, using sample paths to build estimators $\hat f_t \approx \mb{E}[U_{t+1} \vert \mc{F}_t]$. The training algorithm is presented in Algorithm~\ref{alg:alg_stopping_training}.

\begin{algorithm}[t]
	\SetAlgoNoLine
	\KwIn{$N$ sample paths $(H_0^n,H_1^n,\dots,H_T^n)$ for $n \in 0,1,\dots,N-1$.}
	\KwOut{Functions $(\hat f_0,\hat f_1,\dots,\hat f_{T-1})$ that approximate the conditional expectations of the Snell envelope, i.e.$\hat f_t(H_t) \approx \mb{E}[U_{t+1} \vert \mc{F}_t]$.}
	// Initialize the future payoff matrix $CF$ with the base case.\\
	For all $n$, set $CF_T^n = H_T^n$\;
	// Recursively calculate the future payoff matrix $CF$ and $\hat f_t$.\\
	\For{t = T-1,T-2,\dots,0}{
		// Find the sample paths with non-zero payoff at time $t$.\\
		$\mc{N}_t = \{ n : H_t^n > 0 \}$\;
		// Regress the future payoffs on the current value of $H_t$, restricting to sample paths $\mc{N}_t$.\\
		\eIf{$\mc{N}_t$ is empty}{
			$\hat f_t = 0$\;
		}{
			$\hat f_t = $ regress $(\max_{s > t}CF_s^n)_{n \in \mc{N}_t}$ on $(H_t^n)_{n \in \mc{N}_t}$\;
		}
		// Calculate the approximate optimal strategy at time $t$.\\
		\For{all $n$}{
			// This condition means it would better to stop at $t$ than future times on this sample path $n$.\\
			\If{$H_t^n > \hat f_t(H_t^n)$}{
				$CF_t^n = H_t^n$; $CF_{t+1}^n = 0$, $CF_{t+2}^n = 0$, \dots $CF_{T}^n = 0$\;
			}
		}
	}
	\Return{$(\hat f_0,\dots,\hat f_{T-1})$}
	\caption{A recursive method for training the algorithmic stopping method, which approximates the Snell envelope.}
	\label{alg:alg_stopping_training}
\end{algorithm}

Algorithm~\ref{alg:alg_stopping_training} works as follows. It calculates the payoff for each sample path. At time $t$, $(CF_s^n)_{s \ge t}$ represents the future incoming payoff at each time $s \ge t$.
Since the formulation allows us to only stop once, it will be non-zero in at most one time $s$. For emphasis, $CF_s^n$ is non-zero at the time index $s$ where it is optimal to stop given the available information. It will {\em not} choose the maximal element of $H_s^n$, but rather it will choose the first time when the current payoff $H_t^n$ exceeds the estimated future payoffs $\hat f_t(H_t^n)$. 

Many works differ based on the regression method used; we present the algorithm in a fashion that is agnostic to the regression method used. \citet{Carriere:1996aa} uses spline interpolation as a regression technique, whereas~\citet{Tsitsiklis:1999aa} and~\citet{Longstaff:2001aa} choose basis functions and use least-squares regression methods.
The general algorithm is provided in Algorithm~\ref{alg:alg_stopping_training}.

Algorithm~\ref{alg:alg_stopping_training} introduces one relaxation, which is common in the literature~\cite{Longstaff:2001aa}. 
Rather than approximate $\mb{E}[U_{t+1} \vert \mc{F}_t]$, we approximate $\mb{E}[U_{t+1} \vert H_t]$. Note that the space of $\sigma(H_t)$-measurable functions will be a subspace of $\mc{F}_t$-measurable functions (since $H_t \in \mc{F}_t$). This is motivated by computational reasons. For example, if we modeled $\mc{F}_t = \sigma(X_0,\dots,X_t)$ for some random variables $(X_t)_t$, the dimension of the domain of regression would grow across time, requiring increasing amounts of data for longer time horizons. This relaxation is motivated by the intuition that the most recent payoff $H_t$ provides most of the information about future payoffs, i.e. $\mb{E}[U_{t+1} \vert H_t]$ is close to $\mb{E}[U_{t+1} \vert \mc{F}_t]$.

Finally, it is important to note that Algorithm~\ref{alg:alg_stopping_training} has one more quirk: it removes sample paths that currently have zero payoff (i.e. paths $n$ where $H_t^n = 0$) prior to regression. This idea was introduced by~\citet{Longstaff:2001aa}. This is done to reduce bias in the trained estimator, since the process $H_t$ is thresholded at $0$. Put another way, $H_t^n = \pi_t \vee 0$, so the underlying process $(\pi_t)_t$ could be in many different states to yield the payoff $H_t^n = 0$. Thus, when regressing, the data would not be informative when $H_t^n = 0$.
To reduce the bias introduced as a result, we do not include sample paths with zero payoff at time $t$ in the regression.

Finally, once the training is done, to implement the algorithmic stopping method, the deployment is straight-forward. Using the estimators $\hat f_t$, we compare the current payoff $H_t$ with the estimate of the future payoffs $\hat f_t$, and exit when the current payoff is higher. 
Additionally, note that if the decision is to purchase/reject the good at time $t$, this depends only on the values $(H_0,\dots,H_t)$, as desired. Thus, the algorithm defines a valid stopping time. The details are provided in Algorithm~\ref{alg:alg_stopping_implement}.

\begin{algorithm}[t]
	\SetAlgoNoLine
	\KwIn{Trained estimators $(\hat f_t)_t$ and an input stream of payoffs $(H_t)_t$.}
	\KwOut{A time $t$ to purchase/reject the good, and a decision of \texttt{purchase} or \texttt{reject}.}
	\For{t = 0,1,\dots,T-1}{
		// Compare $H_t$ with the estimate of $\mb{E}[U_{t+1} \vert \mc{F}_t]$.\\
		\If{$H_t > \hat f_t(H_t)$}{
			// Make a purchase immediately. \\
			\Return{($t$, \texttt{purchase})}
		}
	}
	// At the final time step, decide whether to purchase or reject based on the payoff.\\
	\eIf{$H_T > 0$}{
		\Return{($T$, \texttt{purchase})}
	}{
		\Return{($T$, \texttt{reject})}
	}
	\caption{Online implementation of the algorithmic stopping method.}
	\label{alg:alg_stopping_implement}
\end{algorithm}

\subsection{Regression via reproducing kernel Hilbert space methods}

Algorithm~\ref{alg:alg_stopping_training} is presented in a fashion that is agnostic to the regression method used. 
For our implementation, we used reproducing kernel Hilbert space (RKHS) methods. 

First, let's present a quick overview on RKHS methods. For more details, we refer the reader to~\citet{Aronszajn:1950aa}. Let $\mc{H}$ be a RKHS whose elements are functions $\hat f$ that map input features $\mc{X}$ to output values in $\mb{R}$. Furthermore, let $k : \mc{X} \times \mc{X} \rightarrow \mb{R}$ be the corresponding kernel of $\mc{H}$. Then, consider the infinite-dimensional, non-parametric problem of minimizing the empirical risk for some dataset $(x_i,y_i)_{i = 1}^d \in (\mc{X} \times \mb{R})^d$:
\begin{equation}
\label{eq:rkhs_infinite_opt}
\min_{\hat f \in \mc{H}} \sum_{i = 1}^d \left\| \hat f(x_i) - y_i \right\|^2
\end{equation}
To find a $\hat f$ that minimizes the above function, the Representer Theorem states that it suffices to search over the span of the finite set of functions $\{x \mapsto k(x_i,x)\}_{i = 1}^d$~\cite{Aronszajn:1950aa}. Thus, the optimization in Equation~\eqref{eq:rkhs_infinite_opt} can be reduced to the finite-dimensional optimization problem:
\[
\min_{w \in \mb{R}^d} \sum_{i = 1}^d \left\| \sum_{j = 1}^d w_j k(x_j,x_i) - y_i \right\|^2
\]
The resulting estimator is determined by $\theta = (w,x_1,\dots,x_d)$:
\[
\hat f_t(x; \theta) = \sum_{j = 1}^d w_j k(x_j,x)
\]
In this paper, we use the Gaussian kernel $k(x,y) = \exp( -\|x-y\|^2 / 2 \sigma^2 )$, parameterized by $\sigma > 0$.

%% file: example.tex

%


In this section, we demonstrate the efficacy of the computational methods introduced in Section~\ref{sec:computation} on a numerical example. For these simulations, we use the model outlined in Section~\ref{sec:example_model}.



\begin{figure}
    \centering
    \includegraphics[width=0.75\columnwidth]{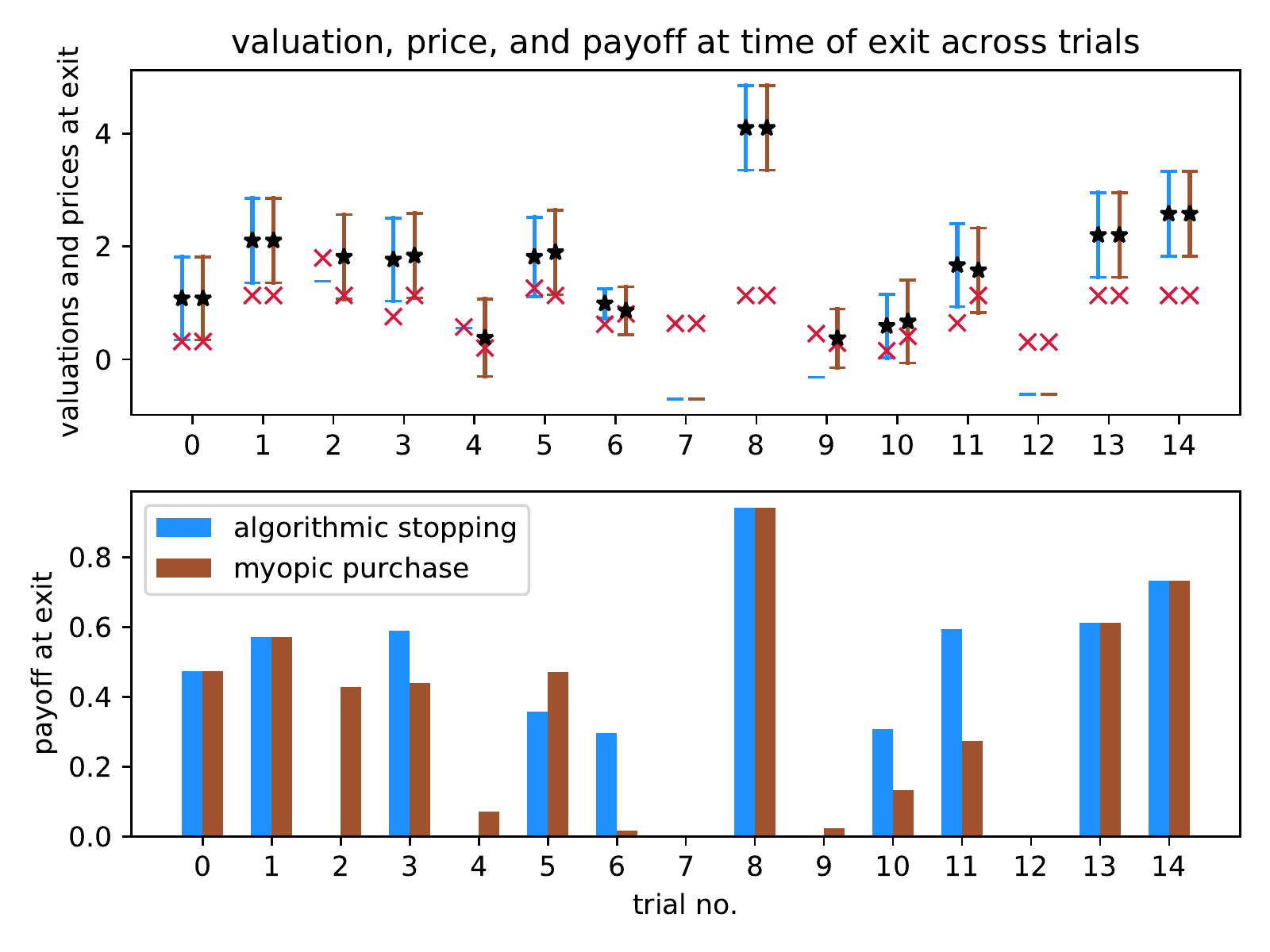}
    \caption{The valuation, price, and payoffs at the time of exit of the two strategies across the same sample paths. In the top plot, the region between the bars represent $\pm$ 1.5 standard deviations from the expected value for the consumer's valuation at the time of exit, chosen by the algorithmic stopping method (light-blue) and the myopic purchasing strategy (brown). The red \texttt{x} indicates the price offered at the time of exit. A black star indicates that a purchase was made, and is placed at the mean of the valuation at the time of exit. The absence of a black star means that no purchase was made. In the bottom plot, the corresponding payoffs from exit are given.}
    \label{fig:trials}
\end{figure}

\begin{figure}
    \centering
    \includegraphics[width=0.75\columnwidth]{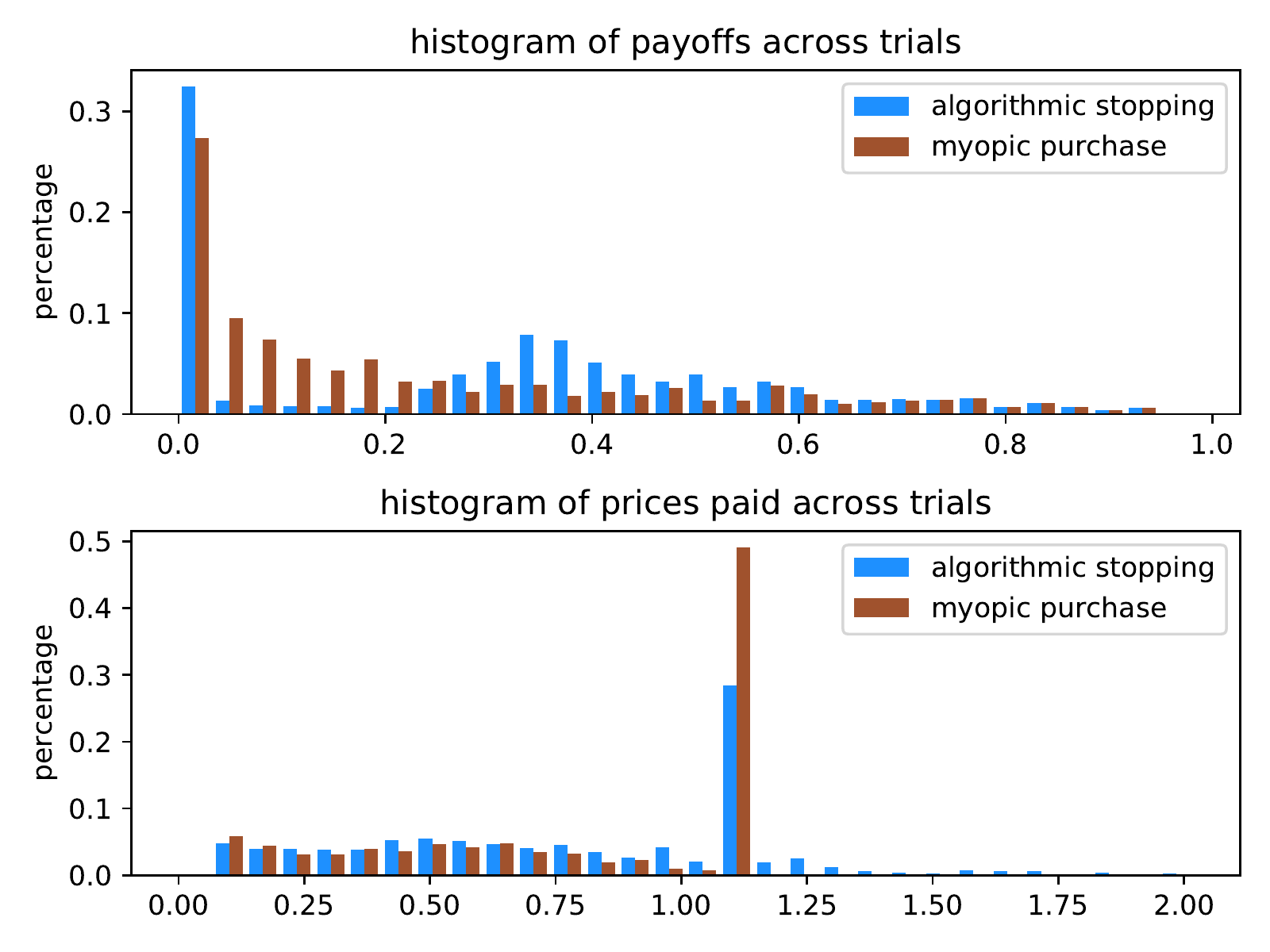}
    \caption{Histograms of the payoffs and prices paid across 1,000 trials. In the top plot, we show the payoffs from following the algorithmic stopping method as well as a myopic purchasing strategy for all 1,000 trials. In the bottom plot, we show the prices paid when following the algorithmic stopping method and myopic purchasing strategy; prices are included only from trials where a purchase was made. Of the 1,000 trials, the algorithmic stopping method made a purchase 685 times and the myopic purchasing strategy made a purchase 825 times.}
    \label{fig:payoff_hist}
\end{figure}

\begin{figure}
    \centering
    \includegraphics[width=0.75\columnwidth]{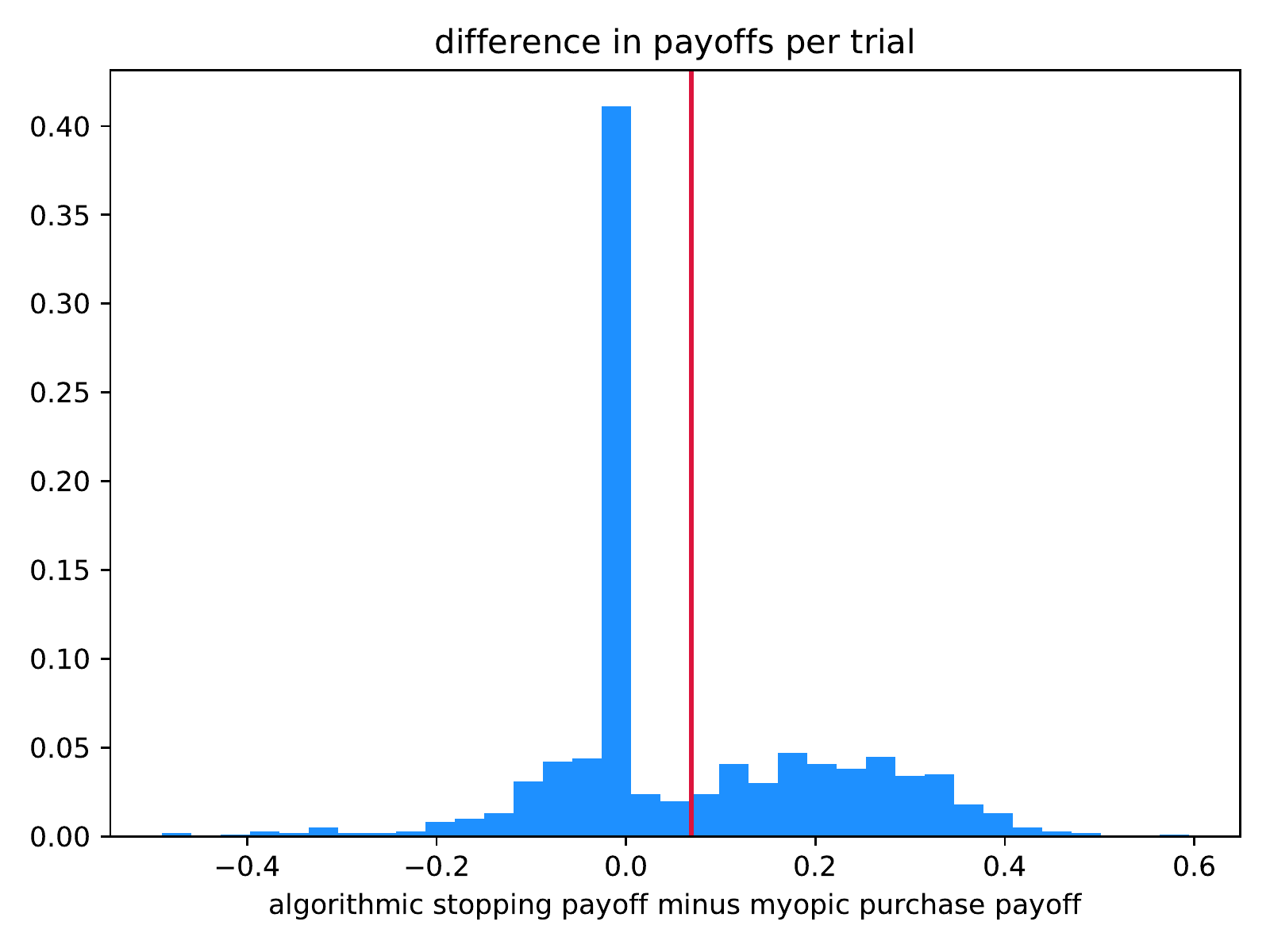}
    \caption{Histograms of the difference between the algorithmic stopping payoff and the myopic purchase payoff across 1,000 trials. The red line indicates the empirical average difference, which is $0.07$.}
    \label{fig:differences}
\end{figure}

We used the following parameters. First, we drew $N = 500$ sample paths as a training dataset for Algorithm~\ref{alg:alg_stopping_training}, which we refer to as the algorithmic stopping method in this section. The time horizon was $T = 25$. The consumer's parameter for risk aversion was $\gamma = 1$. We set $\sigma_\eps = 0.1$, which represents how much the consumer's uncertainty about her own valuation decreases across time. We set $\sigma_\xi = 1$, which is the noise in the seller's observations of the consumer valuation. Similarly, we set the prior for the seller's estimate with the parameters $\mu = 1$ and $\sigma_v = 1$. For the purposes of these simulations, we drew the initial valuations $v_0$ according to the prior $N(\mu,\sigma_v^2)$ as well. Finally, the parameter for the Gaussian kernel in our reproducing kernel Hilbert space method was chosen as $\sigma = 1$. For Figure~\ref{fig:trials}, we used 15 test trials to illustrate the behavior of trajectories, and for Figures~\ref{fig:payoff_hist} and~\ref{fig:differences} we used 1,000 test trials to evaluate the distributions of our algorithm's performance. The data in these figures use estimators trained on $N = 500$ training sample paths.

\begin{figure}
    \centering
    \includegraphics[width=0.75\columnwidth]{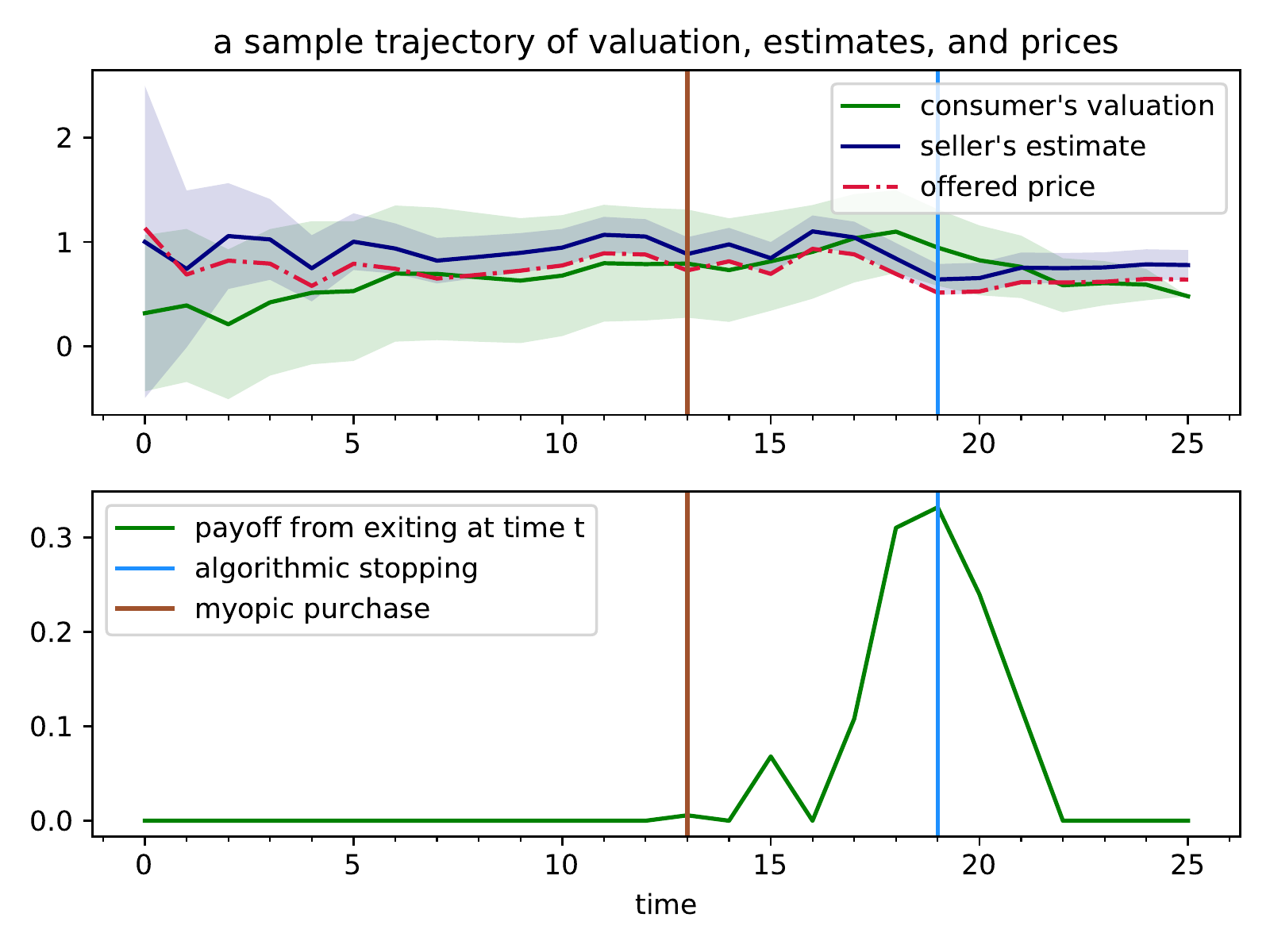}
    \caption{A sample trajectory where the algorithmic stopping method performs better than the myopic purchasing strategy. In the top plot, the green line represents the consumer's uncertainty about her own valuation, the blue line represents the seller's estimate of the consumer's valuation, and the red line represents the prices offered at each time step. The shaded regions represent $\pm 1.5$ standard deviations from the mean. In the bottom plot, the green line represents $H_t = \pi_t \vee 0$, the payoff from exit at each time step. The light-blue line represents the time chosen by the algorithmic stopping method, and the brown line represents the time chosen by a myopic purchasing strategy.}
    \label{fig:sample_traj}
\end{figure}

In Figure~\ref{fig:sample_traj}, we provide a sample trajectory to illustrate the behavior of the algorithmic stopping method at the level of a single sample path. In the top figure, the green regions indicate the consumer's uncertainty about her own valuation decreasing across time as she learns. Similarly, as time passes, the seller makes observations and forms a better estimate of the consumer's willingness to pay, as indicated by the blue line. At each time step, the seller chooses a price to offer the consumer (the red line), which attempts to maximize expected revenue against myopic consumers. For this sample path, we can note a few things. First, even though the expected value of her valuation is above the offered price at time $t = 7$, her payoff is not positive due to the risk aversion in her utility function. However, at time $t = 13$, her payoff is positive due to the decreased uncertainty across time. However, for this realization, the price drops significantly after a few more time steps, and the time chosen by the algorithmic stopping method $t = 19$ yields a significantly higher payoff, due to both the consumer's reduced uncertainty in her own valuation as well as lower offered prices. The algorithmic stopping method decides when it is better in expectation to continue learning, reduce uncertainty, and potentially wait for prices to drop, whereas the myopic purchasing strategy greedily makes a purchase at the first time the payoff is positive.

In Figure~\ref{fig:trials}, we show the valuation, price, and payoff at the time of exit for both the algorithmic stopping method and the myopic purchasing strategy. We see that the behavior along sample paths falls into 3 general categories. 
\begin{itemize}
\item
First, trials 0, 1, 7, 8, 12, 13, and 14 show that the algorithmic stopping method and myopic purchasing strategy agree on their decision of when to exit, and the payoff received is the same as a consequence. 
\item
Second, we see trajectories where the algorithmic stopping method is able to receive higher payoff by waiting for price changes in the future or a reduction in uncertainty, as shown in trials 3, 6, 10, and 11. In trials 3 and 11, the higher payoff are primarily from waiting for lower prices, and in trials 6 and 10, the higher payoff arises primarily from a reduction in uncertainty. 
\item
The third outcome (trials 2, 4, 5, and 9) arises when the algorithmic stopping time decides to wait, and either never sees a good opportunity for purchase again, or eventually has to settle for a higher price than originally offered. In particular, the algorithmic stopping method is not better than the greedy strategy for all sample paths, which is not possible given the limited information available. Rather, it endeavors to be better in expectation. In trials 2, 4, and 9, we see that the algorithmic stopping method never makes a purchase, while the greediness of the myopic purchasing strategy nets some positive payoff. In trial 5, we see that the algorithmic stopping method is forced to make a purchase at a worse price due to bad realizations of the sample path.
\end{itemize}

In Figure~\ref{fig:payoff_hist}, we demonstrate a histogram of payoffs and prices paid across 1,000 trials. In the top figure, we can see that the algorithmic stopping method receives 0 payoff more often than a myopic purchasing strategy; our proposed method encourages waiting for prices to potentially drop in the future, as well as the increased payoff from reduced uncertainty for a risk-averse agent. For some sample paths, it will be better to take the greedy approach and make a purchase as soon as there is positive payoff, which explains why the algorithmic stopping method has a higher percentage of trials at 0. However, in other sample paths, we find that the payoffs are higher for the algorithmic stopping method as a result of its patience, as we can see by the higher weights around payoffs of 0.25 to 0.6. 

From the bottom figure, we can see the distribution of prices paid when purchases are made. The price is only included in this figure if a purchase was made. Since all our trials were initialized to the same seller prior, the offered price at time $t = 0$ is the same for every trial, which explains the giant peak at 1.1, which corresponds to sample paths where a purchase was made immediately. The algorithmic stopping method oftentimes sees lower prices, but also sometimes sees higher prices, as demonstrated in this histogram. This histogram hints that the increased payoffs from the top figure primarily arise due to reduced uncertainty rather than better prices.

In Figure~\ref{fig:differences}, we plot the differences between the payoff chosen by the algorithmic stopping method and the myopic purchase strategy. In 367/1000 trials, both methods choose the same payoff, corresponding a large amount of the mass around $0$. (The histogram shows that about 40\% of sample paths had payoffs very close to each other, even if not exactly equal to each other.) The average increase in payoff from the algorithmic stopping method is $0.07$, which with these simulation parameters, is a significant increase. (For context, the average payoff across 1,000 trials from the myopic purchase strategy is $0.23$.)
We can see that this method is not without downside risk, as well, as there are many sample paths where the payoff is worse. However, we do see improvements in expectation.

%% file: conclusion.tex

In this paper, we formalized the problem a consumer faces when a seller uses personalized pricing algorithms. The consumer must balance between learning more about a potential purchase, and revealing more information about herself in the process. Under the key assumption that the seller credibly commits to pricing strategies, the consumer's decision can be reduced to the selection of a stopping time. Thus, the consumer's problem can be translated into one commonly studied in mathematical finance: the pricing of American options. We present both a theoretical characterization of optimal stopping times as well as algorithms to compute a stopping time given a training dataset of sample paths.

For future work, we hope to better understand pricing strategies: instead of treating pricing strategies as exogenously fixed (as considered in this paper), how would a seller {\it choose} a pricing strategy, given that the consumer is using an algorithmic stopping method to make purchasing decisions? The resulting interaction would become a Stackelberg game, or a bi-level optimization, where the lower-level optimization is an optimization across stopping times. More generally, if we remove the seller's ability to credibly commit to pricing strategies in advance, what are the game-theoretic properties of this interaction? If agents use our algorithmic stopping method, how does this affect the behavior and performance of current personalized pricing algorithms?

Broadly speaking, we believe these types of problems to be understudied. There is significant effort in the study of how to design advertisements and personalize prices to maximize revenue for the sellers in these situations, but these formulations often do not account for the welfare and utility of the consumers they target. In this paper, we focus on algorithmic methods in which users can intelligently decide how much data to share, or when it's best to commit early to certain decisions before some observer learns too much about them. The hope is to develop tools that can protect consumers against the widespread usage of their personal data against them.

%% file: main.bbl

\begin{thebibliography}{32}


\ifx \showCODEN    \undefined \def \showCODEN     #1{\unskip}     \fi
\ifx \showDOI      \undefined \def \showDOI       #1{#1}\fi
\ifx \showISBNx    \undefined \def \showISBNx     #1{\unskip}     \fi
\ifx \showISBNxiii \undefined \def \showISBNxiii  #1{\unskip}     \fi
\ifx \showISSN     \undefined \def \showISSN      #1{\unskip}     \fi
\ifx \showLCCN     \undefined \def \showLCCN      #1{\unskip}     \fi
\ifx \shownote     \undefined \def \shownote      #1{#1}          \fi
\ifx \showarticletitle \undefined \def \showarticletitle #1{#1}   \fi
\ifx \showURL      \undefined \def \showURL       {\relax}        \fi
\providecommand\bibfield[2]{#2}
\providecommand\bibinfo[2]{#2}
\providecommand\natexlab[1]{#1}
\providecommand\showeprint[2][]{arXiv:#2}

\bibitem[\protect\citeauthoryear{Amin, Rostamizadeh, and Syed}{Amin
  et~al\mbox{.}}{2013}]%
        {Amin:2013aa}
\bibfield{author}{\bibinfo{person}{Kareem Amin}, \bibinfo{person}{Afshin
  Rostamizadeh}, {and} \bibinfo{person}{Umar Syed}.}
  \bibinfo{year}{2013}\natexlab{}.
\newblock \showarticletitle{Learning Prices for Repeated Auctions with
  Strategic Buyers}. In \bibinfo{booktitle}{\emph{Proceedings of the 26th
  International Conference on Neural Information Processing Systems - Volume
  1}} (Lake Tahoe, Nevada) \emph{(\bibinfo{series}{NIPS'13})}.
  \bibinfo{publisher}{Curran Associates Inc.}, \bibinfo{address}{Red Hook, NY,
  USA}, \bibinfo{pages}{1169--1177}.
\newblock


\bibitem[\protect\citeauthoryear{Aronszajn}{Aronszajn}{1950}]%
        {Aronszajn:1950aa}
\bibfield{author}{\bibinfo{person}{N. Aronszajn}.}
  \bibinfo{year}{1950}\natexlab{}.
\newblock \showarticletitle{Theory of Reproducing Kernels}.
\newblock \bibinfo{journal}{\emph{Trans. Amer. Math. Soc.}}
  \bibinfo{volume}{68}, \bibinfo{number}{3} (\bibinfo{year}{1950}),
  \bibinfo{pages}{337--404}.
\newblock
\showISSN{00029947}
\urldef\tempurl%
\url{http://www.jstor.org/stable/1990404}
\showURL{%
\tempurl}


\bibitem[\protect\citeauthoryear{Aviv and Pazgal}{Aviv and Pazgal}{2008}]%
        {Aviv:2008aa}
\bibfield{author}{\bibinfo{person}{Yossi Aviv} {and} \bibinfo{person}{Amit
  Pazgal}.} \bibinfo{year}{2008}\natexlab{}.
\newblock \showarticletitle{Optimal Pricing of Seasonal Products in the
  Presence of Forward-Looking Consumers}.
\newblock \bibinfo{journal}{\emph{Manufacturing \& Service Operations
  Management}} \bibinfo{volume}{10}, \bibinfo{number}{3}
  (\bibinfo{year}{2008}), \bibinfo{pages}{339--359}.
\newblock
\urldef\tempurl%
\url{https://doi.org/10.1287/msom.1070.0183}
\showDOI{\tempurl}
\showeprint{https://pubsonline.informs.org/doi/pdf/10.1287/msom.1070.0183}


\bibitem[\protect\citeauthoryear{Ban and Keskin}{Ban and Keskin}{2019}]%
        {Ban:2019aa}
\bibfield{author}{\bibinfo{person}{Gah-Yi Ban} {and} \bibinfo{person}{N.~Bora
  Keskin}.} \bibinfo{year}{2019}\natexlab{}.
\newblock \showarticletitle{Personalized Dynamic Pricing with Machine Learning:
  High Dimensional Features and Heterogeneous Elasticity}.
\newblock \bibinfo{journal}{\emph{SSRN}} (\bibinfo{year}{2019}).
\newblock


\bibitem[\protect\citeauthoryear{Barocas, Hardt, and Narayanan}{Barocas
  et~al\mbox{.}}{2019}]%
        {Barocas:2019aa}
\bibfield{author}{\bibinfo{person}{Solon Barocas}, \bibinfo{person}{Moritz
  Hardt}, {and} \bibinfo{person}{Arvind Narayanan}.}
  \bibinfo{year}{2019}\natexlab{}.
\newblock \bibinfo{booktitle}{\emph{Fairness and Machine Learning}}.
\newblock \bibinfo{publisher}{fairmlbook.org}.
\newblock
\newblock
\shownote{\url{http://www.fairmlbook.org}.}


\bibitem[\protect\citeauthoryear{Besanko and Winston}{Besanko and
  Winston}{1990}]%
        {Besanko:1990aa}
\bibfield{author}{\bibinfo{person}{David Besanko} {and}
  \bibinfo{person}{Wayne~L. Winston}.} \bibinfo{year}{1990}\natexlab{}.
\newblock \showarticletitle{Optimal Price Skimming by a Monopolist Facing
  Rational Consumers}.
\newblock \bibinfo{journal}{\emph{Management Science}} \bibinfo{volume}{36},
  \bibinfo{number}{5} (\bibinfo{year}{1990}), \bibinfo{pages}{555--567}.
\newblock
\urldef\tempurl%
\url{https://doi.org/10.1287/mnsc.36.5.555}
\showDOI{\tempurl}
\showeprint{https://doi.org/10.1287/mnsc.36.5.555}


\bibitem[\protect\citeauthoryear{Carriere}{Carriere}{1996}]%
        {Carriere:1996aa}
\bibfield{author}{\bibinfo{person}{Jacques~F. Carriere}.}
  \bibinfo{year}{1996}\natexlab{}.
\newblock \showarticletitle{Valuation of the early-exercise price for options
  using simulations and nonparametric regression}.
\newblock \bibinfo{journal}{\emph{Insurance: Mathematics and Economics}}
  \bibinfo{volume}{19}, \bibinfo{number}{1} (\bibinfo{year}{1996}),
  \bibinfo{pages}{19 -- 30}.
\newblock
\showISSN{0167-6687}
\urldef\tempurl%
\url{https://doi.org/10.1016/S0167-6687(96)00004-2}
\showDOI{\tempurl}


\bibitem[\protect\citeauthoryear{Chade and de~Serio}{Chade and
  de~Serio}{2002}]%
        {Chade:2002aa}
\bibfield{author}{\bibinfo{person}{Hector Chade} {and}
  \bibinfo{person}{Virginia~Vera de Serio}.} \bibinfo{year}{2002}\natexlab{}.
\newblock \showarticletitle{Pricing, Learning, and Strategic Behavior in a
  Single-Sale Model}.
\newblock \bibinfo{journal}{\emph{Economic Theory}} \bibinfo{volume}{19},
  \bibinfo{number}{2} (\bibinfo{year}{2002}), \bibinfo{pages}{333--353}.
\newblock
\showISSN{09382259, 14320479}
\urldef\tempurl%
\url{http://www.jstor.org/stable/25055471}
\showURL{%
\tempurl}


\bibitem[\protect\citeauthoryear{Cohen, Lobel, and Paes~Leme}{Cohen
  et~al\mbox{.}}{2016}]%
        {Cohen:2016aa}
\bibfield{author}{\bibinfo{person}{Maxime~C. Cohen}, \bibinfo{person}{Ilan
  Lobel}, {and} \bibinfo{person}{Renato Paes~Leme}.}
  \bibinfo{year}{2016}\natexlab{}.
\newblock \showarticletitle{Feature-Based Dynamic Pricing}. In
  \bibinfo{booktitle}{\emph{Proceedings of the 2016 ACM Conference on Economics
  and Computation}} (Maastricht, The Netherlands) \emph{(\bibinfo{series}{EC
  '16})}. \bibinfo{publisher}{Association for Computing Machinery},
  \bibinfo{address}{New York, NY, USA}, \bibinfo{pages}{817}.
\newblock
\showISBNx{9781450339360}
\urldef\tempurl%
\url{https://doi.org/10.1145/2940716.2940728}
\showDOI{\tempurl}


\bibitem[\protect\citeauthoryear{den Boer}{den Boer}{2015}]%
        {Boer:2015aa}
\bibfield{author}{\bibinfo{person}{Arnoud~V. den Boer}.}
  \bibinfo{year}{2015}\natexlab{}.
\newblock \showarticletitle{Dynamic pricing and learning: Historical origins,
  current research, and new directions}.
\newblock \bibinfo{journal}{\emph{Surveys in Operations Research and Management
  Science}} \bibinfo{volume}{20}, \bibinfo{number}{1} (\bibinfo{year}{2015}),
  \bibinfo{pages}{1 -- 18}.
\newblock
\showISSN{1876-7354}
\urldef\tempurl%
\url{https://doi.org/10.1016/j.sorms.2015.03.001}
\showDOI{\tempurl}


\bibitem[\protect\citeauthoryear{F\"{o}llmer and Schied}{F\"{o}llmer and
  Schied}{2008}]%
        {Follmer:2008aa}
\bibfield{author}{\bibinfo{person}{Hans F\"{o}llmer} {and}
  \bibinfo{person}{Alexander Schied}.} \bibinfo{year}{2008}\natexlab{}.
\newblock \bibinfo{booktitle}{\emph{Stochastic Finance: An Introduction in
  Discrete Time} (\bibinfo{edition}{2nd} ed.)}.
\newblock \bibinfo{publisher}{De Gruyter}, \bibinfo{address}{Berlin, Boston}.
\newblock


\bibitem[\protect\citeauthoryear{Golrezaei, Javanmard, and Mirrokni}{Golrezaei
  et~al\mbox{.}}{2019}]%
        {Golrezaei:2019aa}
\bibfield{author}{\bibinfo{person}{Negin Golrezaei}, \bibinfo{person}{Adel
  Javanmard}, {and} \bibinfo{person}{Vahab Mirrokni}.}
  \bibinfo{year}{2019}\natexlab{}.
\newblock \showarticletitle{Dynamic Incentive-aware Learning: Robust Pricing in
  Contextual Auctions}. In \bibinfo{booktitle}{\emph{33rd Conference on Neural
  Information Processing Systems (NeurIPS 2019)}}.
\newblock


\bibitem[\protect\citeauthoryear{Hannak, Soeller, Lazer, Mislove, and
  Wilson}{Hannak et~al\mbox{.}}{2014}]%
        {Hannak:2014aa}
\bibfield{author}{\bibinfo{person}{Aniko Hannak}, \bibinfo{person}{Gary
  Soeller}, \bibinfo{person}{David Lazer}, \bibinfo{person}{Alan Mislove},
  {and} \bibinfo{person}{Christo Wilson}.} \bibinfo{year}{2014}\natexlab{}.
\newblock \showarticletitle{Measuring Price Discrimination and Steering on
  E-Commerce Web Sites}. In \bibinfo{booktitle}{\emph{Proceedings of the 2014
  Conference on Internet Measurement Conference}} (Vancouver, BC, Canada)
  \emph{(\bibinfo{series}{IMC '14})}. \bibinfo{publisher}{Association for
  Computing Machinery}, \bibinfo{address}{New York, NY, USA},
  \bibinfo{pages}{305--318}.
\newblock
\showISBNx{9781450332132}
\urldef\tempurl%
\url{https://doi.org/10.1145/2663716.2663744}
\showDOI{\tempurl}


\bibitem[\protect\citeauthoryear{Hart and Tirole}{Hart and Tirole}{1988}]%
        {Hart:1988aa}
\bibfield{author}{\bibinfo{person}{Oliver~D. Hart} {and} \bibinfo{person}{Jean
  Tirole}.} \bibinfo{year}{1988}\natexlab{}.
\newblock \showarticletitle{Contract Renegotiation and Coasian Dynamics}.
\newblock \bibinfo{journal}{\emph{The Review of Economic Studies}}
  \bibinfo{volume}{55}, \bibinfo{number}{4} (\bibinfo{year}{1988}),
  \bibinfo{pages}{509--540}.
\newblock
\showISSN{00346527, 1467937X}
\urldef\tempurl%
\url{http://www.jstor.org/stable/2297403}
\showURL{%
\tempurl}


\bibitem[\protect\citeauthoryear{Haugh and Kogan}{Haugh and Kogan}{2004}]%
        {Haugh:2004aa}
\bibfield{author}{\bibinfo{person}{Martin~B. Haugh} {and}
  \bibinfo{person}{Leonid Kogan}.} \bibinfo{year}{2004}\natexlab{}.
\newblock \showarticletitle{Pricing American Options: A Duality Approach}.
\newblock \bibinfo{journal}{\emph{Operations Research}} \bibinfo{volume}{52},
  \bibinfo{number}{2} (\bibinfo{year}{2004}), \bibinfo{pages}{258--270}.
\newblock
\showISSN{0030364X, 15265463}
\urldef\tempurl%
\url{http://www.jstor.org/stable/30036577}
\showURL{%
\tempurl}


\bibitem[\protect\citeauthoryear{Javanmard and Nazerzadeh}{Javanmard and
  Nazerzadeh}{2019}]%
        {Javanmard:2019aa}
\bibfield{author}{\bibinfo{person}{Adel Javanmard} {and} \bibinfo{person}{Hamid
  Nazerzadeh}.} \bibinfo{year}{2019}\natexlab{}.
\newblock \showarticletitle{Dynamic Pricing in High-Dimensions}.
\newblock \bibinfo{journal}{\emph{J. Mach. Learn. Res.}} \bibinfo{volume}{20},
  \bibinfo{number}{1} (\bibinfo{date}{Jan.} \bibinfo{year}{2019}),
  \bibinfo{pages}{315--363}.
\newblock
\showISSN{1532-4435}


\bibitem[\protect\citeauthoryear{Kanoria and Nazerzadeh}{Kanoria and
  Nazerzadeh}{2014}]%
        {Kanoria:2014aa}
\bibfield{author}{\bibinfo{person}{Yash Kanoria} {and} \bibinfo{person}{Hamid
  Nazerzadeh}.} \bibinfo{year}{2014}\natexlab{}.
\newblock \showarticletitle{Dynamic Reserve Prices for Repeated Auctions:
  Learning from Bids}. In \bibinfo{booktitle}{\emph{Web and Internet
  Economics}}, \bibfield{editor}{\bibinfo{person}{Tie-Yan Liu},
  \bibinfo{person}{Qi~Qi}, {and} \bibinfo{person}{Yinyu Ye}} (Eds.).
  \bibinfo{publisher}{Springer International Publishing},
  \bibinfo{address}{Cham}, \bibinfo{pages}{232--232}.
\newblock
\showISBNx{978-3-319-13129-0}


\bibitem[\protect\citeauthoryear{Levin, McGill, and Nediak}{Levin
  et~al\mbox{.}}{2010}]%
        {Levin:2010ab}
\bibfield{author}{\bibinfo{person}{Yuri Levin}, \bibinfo{person}{Jeff McGill},
  {and} \bibinfo{person}{Mikhail Nediak}.} \bibinfo{year}{2010}\natexlab{}.
\newblock \showarticletitle{Optimal Dynamic Pricing of Perishable Items by a
  Monopolist Facing Strategic Consumers}.
\newblock \bibinfo{journal}{\emph{Production and Operations Management}}
  \bibinfo{volume}{19}, \bibinfo{number}{1} (\bibinfo{year}{2010}),
  \bibinfo{pages}{40--60}.
\newblock
\urldef\tempurl%
\url{https://doi.org/10.1111/j.1937-5956.2009.01046.x}
\showDOI{\tempurl}
\showeprint{https://onlinelibrary.wiley.com/doi/pdf/10.1111/j.1937-5956.2009.01046.x}


\bibitem[\protect\citeauthoryear{Levina, Levin, McGill, and Nediak}{Levina
  et~al\mbox{.}}{2009}]%
        {Levina:2009aa}
\bibfield{author}{\bibinfo{person}{Tatsiana Levina}, \bibinfo{person}{Yuri
  Levin}, \bibinfo{person}{Jeff McGill}, {and} \bibinfo{person}{Mikhail
  Nediak}.} \bibinfo{year}{2009}\natexlab{}.
\newblock \showarticletitle{Dynamic Pricing with Online Learning and Strategic
  Consumers: An Application of the Aggregating Algorithm}.
\newblock \bibinfo{journal}{\emph{Operations Research}} \bibinfo{volume}{57},
  \bibinfo{number}{2} (\bibinfo{year}{2009}), \bibinfo{pages}{327--341}.
\newblock
\urldef\tempurl%
\url{https://doi.org/10.1287/opre.1080.0577}
\showDOI{\tempurl}
\showeprint{https://doi.org/10.1287/opre.1080.0577}


\bibitem[\protect\citeauthoryear{Loginova and Taylor}{Loginova and
  Taylor}{2008}]%
        {Loginova:2008aa}
\bibfield{author}{\bibinfo{person}{Oksana Loginova} {and}
  \bibinfo{person}{Curtis~R. Taylor}.} \bibinfo{year}{2008}\natexlab{}.
\newblock \showarticletitle{Price experimentation with strategic buyers}.
\newblock \bibinfo{journal}{\emph{Review of Economic Design}}
  \bibinfo{volume}{12}, \bibinfo{number}{3} (\bibinfo{date}{01 Sep}
  \bibinfo{year}{2008}), \bibinfo{pages}{165--187}.
\newblock
\showISSN{1434-4750}
\urldef\tempurl%
\url{https://doi.org/10.1007/s10058-008-0048-5}
\showDOI{\tempurl}


\bibitem[\protect\citeauthoryear{Longstaff and Schwartz}{Longstaff and
  Schwartz}{2001}]%
        {Longstaff:2001aa}
\bibfield{author}{\bibinfo{person}{Francis~A. Longstaff} {and}
  \bibinfo{person}{Eduardo~S. Schwartz}.} \bibinfo{year}{2001}\natexlab{}.
\newblock \showarticletitle{Valuing American Options by Simulation: A Simple
  Least-Squares Approach}.
\newblock \bibinfo{journal}{\emph{The Review of Financial Studies}}
  \bibinfo{volume}{14}, \bibinfo{number}{1} (\bibinfo{year}{2001}),
  \bibinfo{pages}{113--147}.
\newblock
\showISSN{08939454, 14657368}
\urldef\tempurl%
\url{http://www.jstor.org/stable/2696758}
\showURL{%
\tempurl}


\bibitem[\protect\citeauthoryear{Mao, Leme, and Schneider}{Mao
  et~al\mbox{.}}{2018}]%
        {Mao:2018aa}
\bibfield{author}{\bibinfo{person}{Jieming Mao}, \bibinfo{person}{Renato Leme},
  {and} \bibinfo{person}{Jon Schneider}.} \bibinfo{year}{2018}\natexlab{}.
\newblock \showarticletitle{Contextual Pricing for Lipschitz Buyers}.
\newblock In \bibinfo{booktitle}{\emph{Advances in Neural Information
  Processing Systems 31}}, \bibfield{editor}{\bibinfo{person}{S.~Bengio},
  \bibinfo{person}{H.~Wallach}, \bibinfo{person}{H.~Larochelle},
  \bibinfo{person}{K.~Grauman}, \bibinfo{person}{N.~Cesa-Bianchi}, {and}
  \bibinfo{person}{R.~Garnett}} (Eds.). \bibinfo{publisher}{Curran Associates,
  Inc.}, \bibinfo{pages}{5643--5651}.
\newblock
\urldef\tempurl%
\url{http://papers.nips.cc/paper/7807-contextual-pricing-for-lipschitz-buyers.pdf}
\showURL{%
\tempurl}


\bibitem[\protect\citeauthoryear{Nambiar, Simchi-Levi, and Wang}{Nambiar
  et~al\mbox{.}}{2018}]%
        {Nambiar:2018aa}
\bibfield{author}{\bibinfo{person}{Mila Nambiar}, \bibinfo{person}{David
  Simchi-Levi}, {and} \bibinfo{person}{He Wang}.}
  \bibinfo{year}{2018}\natexlab{}.
\newblock \showarticletitle{Dynamic Learning and Pricing with Model
  Misspecification}.
\newblock \bibinfo{journal}{\emph{SSRN}} (\bibinfo{year}{2018}).
\newblock


\bibitem[\protect\citeauthoryear{{Qiang} and {Bayati}}{{Qiang} and
  {Bayati}}{2016}]%
        {Qiang:2016aa}
\bibfield{author}{\bibinfo{person}{Sheng {Qiang}} {and} \bibinfo{person}{Mohsen
  {Bayati}}.} \bibinfo{year}{2016}\natexlab{}.
\newblock \showarticletitle{{Dynamic Pricing with Demand Covariates}}.
\newblock \bibinfo{journal}{\emph{arXiv e-prints}}, Article
  \bibinfo{articleno}{arXiv:1604.07463} (\bibinfo{date}{Apr}
  \bibinfo{year}{2016}).
\newblock
\showeprint[arxiv]{stat.ML/1604.07463}


\bibitem[\protect\citeauthoryear{Rogers}{Rogers}{2002}]%
        {Rogers:2002aa}
\bibfield{author}{\bibinfo{person}{L.~C.~G. Rogers}.}
  \bibinfo{year}{2002}\natexlab{}.
\newblock \showarticletitle{Monte Carlo valuation of American options}.
\newblock \bibinfo{journal}{\emph{Mathematical Finance}} \bibinfo{volume}{12},
  \bibinfo{number}{3} (\bibinfo{year}{2002}), \bibinfo{pages}{271--286}.
\newblock
\urldef\tempurl%
\url{https://doi.org/10.1111/1467-9965.02010}
\showDOI{\tempurl}
\showeprint{https://onlinelibrary.wiley.com/doi/pdf/10.1111/1467-9965.02010}


\bibitem[\protect\citeauthoryear{Salant}{Salant}{1989}]%
        {Salant:1989aa}
\bibfield{author}{\bibinfo{person}{Stephen~W. Salant}.}
  \bibinfo{year}{1989}\natexlab{}.
\newblock \showarticletitle{When is Inducing Self-Selection Suboptimal For a
  Monopolist?}
\newblock \bibinfo{journal}{\emph{The Quarterly Journal of Economics}}
  \bibinfo{volume}{104}, \bibinfo{number}{2} (\bibinfo{year}{1989}),
  \bibinfo{pages}{391--397}.
\newblock
\showISSN{00335533, 15314650}
\urldef\tempurl%
\url{http://www.jstor.org/stable/2937854}
\showURL{%
\tempurl}


\bibitem[\protect\citeauthoryear{{Shah}, {Blanchet}, and {Johari}}{{Shah}
  et~al\mbox{.}}{2019}]%
        {Shah:2019aa}
\bibfield{author}{\bibinfo{person}{Virag {Shah}}, \bibinfo{person}{Jose
  {Blanchet}}, {and} \bibinfo{person}{Ramesh {Johari}}.}
  \bibinfo{year}{2019}\natexlab{}.
\newblock \showarticletitle{{Semi-parametric dynamic contextual pricing}}.
\newblock \bibinfo{journal}{\emph{arXiv e-prints}}, Article
  \bibinfo{articleno}{arXiv:1901.02045} (\bibinfo{date}{Jan}
  \bibinfo{year}{2019}).
\newblock
\showeprint[arxiv]{cs.LG/1901.02045}


\bibitem[\protect\citeauthoryear{Shukla, Kolbeinsson, Otwell, Marla, and
  Yellepeddi}{Shukla et~al\mbox{.}}{2019}]%
        {Shukla:2019aa}
\bibfield{author}{\bibinfo{person}{Naman Shukla},
  \bibinfo{person}{Arinbj\"{o}rn Kolbeinsson}, \bibinfo{person}{Ken Otwell},
  \bibinfo{person}{Lavanya Marla}, {and} \bibinfo{person}{Kartik Yellepeddi}.}
  \bibinfo{year}{2019}\natexlab{}.
\newblock \showarticletitle{Dynamic Pricing for Airline Ancillaries with
  Customer Context}. In \bibinfo{booktitle}{\emph{Proceedings of the 25th ACM
  SIGKDD International Conference on Knowledge Discovery and Data Mining}}
  (Anchorage, AK, USA) \emph{(\bibinfo{series}{KDD '19})}.
  \bibinfo{publisher}{Association for Computing Machinery},
  \bibinfo{address}{New York, NY, USA}, \bibinfo{pages}{2174--2182}.
\newblock
\showISBNx{9781450362016}
\urldef\tempurl%
\url{https://doi.org/10.1145/3292500.3330746}
\showDOI{\tempurl}


\bibitem[\protect\citeauthoryear{Stokey}{Stokey}{1979}]%
        {Stokey:1979aa}
\bibfield{author}{\bibinfo{person}{Nancy~L. Stokey}.}
  \bibinfo{year}{1979}\natexlab{}.
\newblock \showarticletitle{Intertemporal Price Discrimination}.
\newblock \bibinfo{journal}{\emph{The Quarterly Journal of Economics}}
  \bibinfo{volume}{93}, \bibinfo{number}{3} (\bibinfo{year}{1979}),
  \bibinfo{pages}{355--371}.
\newblock
\showISSN{00335533, 15314650}
\urldef\tempurl%
\url{http://www.jstor.org/stable/1883163}
\showURL{%
\tempurl}


\bibitem[\protect\citeauthoryear{Su}{Su}{2007}]%
        {Su:2007aa}
\bibfield{author}{\bibinfo{person}{Xuanming Su}.}
  \bibinfo{year}{2007}\natexlab{}.
\newblock \showarticletitle{Intertemporal Pricing with Strategic Customer
  Behavior}.
\newblock \bibinfo{journal}{\emph{Management Science}} \bibinfo{volume}{53},
  \bibinfo{number}{5} (\bibinfo{year}{2007}), \bibinfo{pages}{726--741}.
\newblock
\urldef\tempurl%
\url{https://doi.org/10.1287/mnsc.1060.0667}
\showDOI{\tempurl}
\showeprint{https://doi.org/10.1287/mnsc.1060.0667}


\bibitem[\protect\citeauthoryear{{Tsitsiklis} and {van Roy}}{{Tsitsiklis} and
  {van Roy}}{1999}]%
        {Tsitsiklis:1999aa}
\bibfield{author}{\bibinfo{person}{J.~N. {Tsitsiklis}} {and}
  \bibinfo{person}{B. {van Roy}}.} \bibinfo{year}{1999}\natexlab{}.
\newblock \showarticletitle{Optimal stopping of Markov processes: Hilbert space
  theory, approximation algorithms, and an application to pricing
  high-dimensional financial derivatives}.
\newblock \bibinfo{journal}{\emph{IEEE Trans. Automat. Control}}
  \bibinfo{volume}{44}, \bibinfo{number}{10} (\bibinfo{date}{Oct}
  \bibinfo{year}{1999}), \bibinfo{pages}{1840--1851}.
\newblock
\showISSN{2334-3303}
\urldef\tempurl%
\url{https://doi.org/10.1109/9.793723}
\showDOI{\tempurl}


\bibitem[\protect\citeauthoryear{Varian}{Varian}{1985}]%
        {Varian:1985aa}
\bibfield{author}{\bibinfo{person}{Hal~R. Varian}.}
  \bibinfo{year}{1985}\natexlab{}.
\newblock \showarticletitle{Price Discrimination and Social Welfare}.
\newblock \bibinfo{journal}{\emph{The American Economic Review}}
  \bibinfo{volume}{75}, \bibinfo{number}{4} (\bibinfo{year}{1985}),
  \bibinfo{pages}{870--875}.
\newblock
\showISSN{00028282}
\urldef\tempurl%
\url{http://www.jstor.org/stable/1821366}
\showURL{%
\tempurl}


\end{thebibliography}
